\documentclass[print]{aa}
\usepackage{graphicx}
\usepackage{txfonts}
\usepackage{longtable, lscape}
\usepackage{verbatim}
\usepackage{natbib}
\bibpunct{(}{)}{;}{a}{}{,} 

\begin{document}
   \title{Kinematic parameters and membership probabilities of open clusters in the Bordeaux PM2000 catalogue\thanks{The Tables 1, 4, and 5 are available in electronic form at the CDS via anonymous ftp to {\tt cdsarc.u-strasbg.fr (130.79.128.5)} or via {\tt http://cdsweb.u-strasbg.fr/cgi-bin/qcat?J/A+A/}}}

   \author{A. Krone-Martins
          \inst{1,2}
          ,
          C. Soubiran\inst{2}
          ,
          C. Ducourant\inst{2,1}
          ,
          R. Teixeira\inst{1,2}
          \and
          \\
          J. F. Le Campion\inst{2}          
          }

   \offprints{algol@astro.iag.usp.br}

   \institute{Instituto de Astronomia, Geof\'isica e Ci\^encias Atmosf\'ericas, 
               Universidade de S\~ao Paulo, Rua do Mat\~ao, 1226, Cidade Universit\'aria,
               05508-900 S\~ao Paulo-SP, Brazil
         \and
             Observatoire Aquitain des Sciences de l'Univers, Laboratoire d'Astrophysique de
             Bordeaux, CNRS-UMR 5804, BP 89, 33271 Floirac Cedex, France\\
             }

   \date{Received 16 December 2009 / Accepted 16 February 2010}
 
  \abstract
   {}
   {We derive lists of proper-motions and kinematic membership probabilities for 49 open clusters and possible open clusters in the zone of the Bordeaux PM2000 proper motion catalogue ($+11^{\circ}\le\delta\le+18^{\circ}$). We test different parametrisations of the proper motion and position distribution functions and select the most successful one. In the light of those results, we analyse some objects individually.}
   {The segregation between cluster and field member stars, and the assignment of membership probabilities, is accomplished by applying a new and fully automated method based on both parametrisations of the proper motion and position distribution functions, and genetic algorithm optimization heuristics associated with a derivative-based hill climbing algorithm for the likelihood optimization.}
   {We present a catalogue comprising kinematic parameters and associated membership probability lists for 49 open clusters and possible open clusters in the Bordeaux PM2000 catalogue region. We note that this is the first determination of proper motions for five open clusters. We confirm the non-existence of two kinematic populations in the region of 15 previously suspected non-existent objects.}
   {}


\titlerunning{Kinematics and membership of PM2000 zone open clusters}
\authorrunning{Krone-Martins et al.}

\maketitle


\section{Introduction}
Once the problem of selecting their physical members is resolved, open clusters are widely respected to be a most valuable tool for undertaking studies of our Galaxy and stellar astrophysics. These objects have been used, for example, to determine the spiral structure of the Galaxy and investigate star formation and evolution processes. They are particularly important as tracers of the dynamics \citep{Frinchaboy2008} and the chemical evolution of our Galaxy's disk \citep{Friel1995}. In the advent of high precision astrophysical surveys such as Gaia \citep{Gaia}, their contribution to astrophysical studies should become increasingly important.

However, to establish a coherent understanding of our Galaxy, one needs to use a significant number of open clusters with consistently measured astrophysical parameters, and there have been numerous efforts in this direction \citep{Kharchenko2005, Bragaglia2006, Frinchaboy2008}. However, as noted by \cite{Frinchaboy2008}, the inaccuracy of current membership determinations poses difficulties in conducting these studies on a large scale.

The main advantage of using open clusters in these studies is that once the complex problem of stellar membership is resolved one can derive their main physical parameters such as distance, age, and metallicity. This membership determination is traditionally performed using the stellar kinematics, but in principle one could use a multidimensional space, using for example, spatial and kinematic information, as in \cite{ZhaoChenWen06}, or CMD-isochrone information, as in \cite{Kharchenko2005}. Nonetheless, we argue that, when analysing these objects, one should rely more on kinematics and as little as possible on a single age CMD-based model analysis as in the aforementioned study. This is because, as it has been reported in the literature, the star formation in some open clusters could be non-coeval, as seems to be true in NGC 3603 \citep{Eisenhauer98} and 14 other open clusters \citep{Strobel92}. An indication of an abundance spread has been reported for one of these objects \citep{Frinchaboy08b}.

Based on these results, we undertook a purely kinematic determination of open cluster membership probabilities for objects located in the zone covered by the high precision proper motion catalogue PM2000 \citep{Ducourant2006}. We used a fully automated optimization method and a set of modified parametrisations for the probability distribution functions based on Zhao et al. (1990, 2006). 

This paper is organised as follows. In Sect. 2, we describe the data we used and its selection process. In Sect. 3, we present the methodology and algorithms chosen to obtain the membership lists and the cluster kinematic parameters. In Sect. 4, we describe the validation of the method. In Sect. 5 we present our results, and our comment on some individual objects. Finally, in Sect. 6 we present the conclusions of our current study.

\section{Data}
\subsection{The PM2000 catalogue}
The PM2000 catalogue \citep{Ducourant2006} is a proper motion catalogue that comprises about 2.6 million stars in the declination zone $+11^{\circ}\le\delta\le+18^{\circ}$ and contains positions and proper motions on the ICRS (International Celestial Reference System), as well as meridian magnitudes $V_M$. It was derived from the compilation of systematic drift-scan observations in the Bordeaux Carte du Ciel Zone with the Bordeaux automated meridian circle \citep{Viateau1999} carried out over four years, the reduction of 512 {\it Carte du Ciel} plates (epoch $t\approx1900$) of the Bordeaux zone \citep{Rapaport2001} scanned at the APM Cambridge, and the catalogues AC2000.2 ($t\approx1907$), USNO-A2.0 ($t\approx1950$) and the unpublished USNO Yellow Sky (YS3, $t\approx1978$). The positional precision ranges from 50 to 70 mas, while the proper-motion precision varies from 1.5 to 6 mas yr$^{-1}$, depending on the magnitude. All the data was analysed using a global iterative astrometric reduction \citep{Teixeira1992, Benevides1992, Ducourant1991}.

The catalogue is complete to $V_M=15.4$, with a limiting magnitude of $V_M=16.2$, and typical error of $0.03$ mag ($9.5\le V_M \le13.5$). In addition, a cross identification between all sources in the PM2000 and 2MASS \citep{Cutri2003} was performed, so the PM2000 catalogue also includes 2MASS photometry information for its objects.

\subsection{The clusters sample}
The starting point of our analysis is a list of $49$ open clusters inside the PM2000 declination zone found in the D07 catalogue \citep{Dias07}. We performed a visual inspection of all these clusters by using the Aladin Sky Atlas \citep{Aladin} to verify their coordinates. During this visual check, we noticed that the clusters Berkeley 29, 43, 45, and 47 needed to be slightly recentered about 3'25", 3'39", 1'13", and 1'27", respectively.

The objects were then separated into four different classes: $A$-{\it reference}, $B$-{\it known}, $C$-{\it known without proper-motion determination}, and $D$-{\it others} (including doubtful objects). An object was classified as $D$ and not $C$ when one or more of the following conditions applied:

   \begin{enumerate}
	\item No entry in the WEBDA database;

	\item Classification as {\it not found}, {\it dubious}, {\it no cluster}, or {\it non-existent NGC} in the D07 catalogue;

	\item Existence of some study that excludes the physical clustering of its stars (such as for NGC 1807). 
   \end{enumerate}

For each cluster, we extracted from the PM2000 catalogue all the stars inside a circular area around the apparent cluster centre. We multiplied the cluster diameter by 1.5, to include most of the cluster members in the extracted zones. Two arcminutes were then added to ensure that even for a small cluster, located in a poorly populated region of the Galaxy, a sufficient number of field stars would be available in the extracted zone to enable us to discriminate between the cluster and the field (as in the case of NGC 7772). We visually inspected the DSS images of all the extracted zones, to ensure that the objects were inside the extracted regions if they were clearly identified. We note that this Òextended regionÓ could add some noise (in the form of field stars) to a small object located in a densely populated region. All the extractions were performed by using automated PERL scripts and CDS's vizquery tool. Table \ref{table:1} shows the input list of clusters that we used, as well as the class that we assigned to each object.

The data used to obtain the cluster kinematical parameters were selected from the extracted data by rejecting  on the basis of proper motion errors ($\epsilon_\mu \le 6.0$ mas yr$^{-1}$) and the total proper motion ($|\mu| < 30.0$ mas yr$^{-1}$). This was justified since high $\mu$ field stars cause the proper motion distribution to be flattened, as previously noted in the literature \citep[][and references therein]{BalaguerNunez2004}. Nonetheless, we computed membership probabilities for all the stars in the extracted zones.

\begin{table*}[p]
\caption{Our cluster sample sorted by class type. The parameter values are taken from the D07 catalogue at Vizier \citep{Dias07}, apart from the coordinates of the clusters Berkeley 29, 43, 45, and 47 since they were re-centred by means of a detailed visual examination performed with Aladin (*). The Galactic coordinates presented here were computed by VizieR. Classes:  $A$-{\it reference}, $B$-{\it known}, $C$-{\it known without proper-motion} and $D$-{\it others}. In the remarks, pocr indicates the possible open cluster remnants while the ``c." indicates which condition of the section 2.2 applies. Note: the Chupina clusters are actually subclusters formed on the outskirts of NGC 2682 \citep{Chupina1998}.} 
\label{table:1}
\centering
\begin{tabular}{l l r r r r r r r}
\hline\hline
Cluster & IAU number  & $\alpha_{J2000}$ & $\delta_{J2000}$ & $l(^{\circ})$ & $b(^{\circ})$ & Diam (') & Class & Remarks\\
\hline                        
NGC 2682       & C 0847+120 & 08:51:18        & +11:48:00    & 215.696	& 31.896      & 25.0 & $A$ & \\
\hline                         
NGC 1663       & C 0445+130 & 04:48:58        & +13:08:54  	& 185.845	& -19.735      & 12.0 & $B$ & pocr \\
NGC 1817       & C 0509+166 & 05:12:15        & +16:41:24    & 186.156	& -13.096      & 16.0 & $B$ & \\
NGC 2169       & C 0605+139 & 06:08:24        & +13:57:54    & 195.608	& -2.935      &  5.0 & $B$ & \\
NGC 2194       & C 0611+128 & 06:13:45        & +12:48:24    & 197.250	& -2.350      &  9.0 & $B$ & \\
Berkeley 29*    & C 0650+169 & 06:53:04        & +16:55:41    & 197.948	& 7.980      &  6.0 & $B$ & \\
NGC 2304       & C 0652+180 & 06:55:11        & +17:59:18    & 197.207	& 8.897      &  3.0 & $B$ & \\
NGC 2355       & C 0714+138 & 07:16:59        & +13:45:00    & 203.390	& 11.803      &  7.0 & $B$ & \\
NGC 2395       & C 0724+136 & 07:27:12        & +13:36:30    & 204.605	& 13.988      & 14.0 & $B$ & \\
Chupina 1       &          & 08:50:07        & +11:56:42    & 215.399	& 31.694      &  5.0 & $B$ & \\
Chupina 2       &          & 08:50:30        & +12:17:42    & 215.070	& 31.925      &  2.1 & $B$ & \\
Chupina 3       &          & 08:51:27        & +11:23:42    & 216.147	& 31.759      &  2.9 & $B$ & \\
Chupina 4       &          & 08:52:00        & +12:22:42    & 215.159	& 32.293      &  5.0 & $B$ & \\
Chupina 5       &          & 08:53:01        & +11:51:42    & 215.837	& 32.304      &  4.5 & $B$ & \\
Berkeley 82     & C 1909+129 & 19:11:20        & +13:07:06    & 46.853	& 1.624      &  2.0 & $B$ & \\
Roslund 1       & C 1942+174 & 19:45:00        & +17:31:00    &  54.600	& -3.400      &  3.0 & $B$ & \\
NGC 7036       &          & 21:10:02        & +15:31:00    & 64.544	& -21.443      &  4.0 & $B$ & pocr \\
\hline                          
Skiff J0614+129  &          & 06:14:48        & +12:52:24    & 197.314	& -2.098      &  7.0 & $C$ & \\
Ivanov 2        &          & 06:15:53        & +14:16:00    & 196.213	& -1.198      &  2.0 & $C$ & \\
Berkeley 43*    & C 1913+111 & 19:15:32        & +11:16:31    & 45.696	& -0.140      &  5.0 & $C$ & \\
Berkeley 45*    & C 1916+156 & 19:19:07        & +15:43:07    & 50.033	& 1.146      &  2.0 & $C$ & \\
Alessi 57       &          & 19:20:54        & +15:40:36    & 50.197	& 0.765      &  2.5 & $C$ & \\
Berkeley 47*    & C 1926+173 & 19:28:30        & +17:21:52    & 52.546	& -0.039      &  3.0 & $C$ & \\
King 26         & C 1926+147 & 19:29:00        & +14:52:00    & 50.410	& -1.339      &  2.0 & $C$ & \\
Dias 8          &          & 19:52:07        & +11:37:54    & 50.335	& -7.826      &  2.3 & $C$ & \\
NGC 7772       & C 2349+159 & 23:51:46        & +16:14:48    & 102.739	& -44.273      &  3.0 & $C$ & pocr\\
\hline                          
NGC 1807       & C 0507+164 & 05:10:43        & +16:31:18    & 186.088	& -13.495      & 15.0 & $D$ & (c.3)\\
DolDzim 2       &          & 05:23:54        & +11:28:00    & 192.240	& -13.560      & 10.0 & $D$ & (c.2)\\
Teutsch 11      &          & 06:25:24        & +13:51:59    & 197.654	& 0.649      &  2.3 & $D$ & (c.1) \\
Teutsch 12      &          & 06:25:40        & +13:36:25    & 197.914	& 0.585      &  4.2 & $D$ & (c.1)\\
NGC 2224       &          & 06:27:32        & +12:39:20    & 198.968	& 0.544      & 20.0 & $D$ & (c.2)\\
NGC 2234       &          & 06:29:29        & +16:43:08    & 195.584	& 2.846      & 35.0 & $D$ & (c.2)\\
NGC 2265       &          & 06:41:41        & +11:54:18    & 201.225	& 3.268      &  9.0 & $D$ & (c.2)\\
Dolidze 26      & C 0727+120 & 07:30:06        & +11:54:00    & 206.508	& 13.901      & 23.0 & $D$ & (c.2)\\
NGC 2678       &          & 08:50:02        & +11:20:18    & 216.035	& 31.421      & 10.0 & $D$ & (c.1)\\
DolDzim 7       &          & 17:10:36        & +15:32:00    & 36.294	& 29.166      &  5.0 & $D$ & (c.2)\\
NGC 6525       &          & 18:02:06        & +11:01:24    & 37.378	& 15.890      &  8.0 & $D$ & (c.2)\\
NGC 6738       & C 1859+115 & 19:01:21        & +11:36:54    & 44.398	& 3.102      & 15.0 & $D$ & (c.2) \\
Riddle 15       &          & 19:11:09        & +14:50:04    & 48.357	& 2.455      &  0.8 & $D$ & (c.1)\\
Juchert 1       &          & 19:22:32        & +12:40:00    & 47.727	& -1.001      &  3.2 & $D$ & (c.1)\\
Kronberger 13   &          & 19:25:15        & +13:56:42    & 49.167	& -0.979      &  1.5 & $D$ & (c.1)\\
Dolidze 35      & C 1924+115 & 19:25:24        & +11:39:30    & 47.170	& -2.095      &  7.0 & $D$ & (c.2)\\
NGC 6837       & C 1951+115 & 19:53:08        & +11:41:54    & 50.519	& -8.009	   &  3.0 & $D$ & (c.2)\\
NGC 6839       &          & 19:54:33        & +17:56:18    & 56.114	& -5.152      &  6.0 & $D$ & (c.2)\\
NGC 6840       &          & 19:55:18        & +12:07:36    & 51.162	& -8.253	   &  6.0 & $D$ & (c.2)\\
NGC 6843       &          & 19:56:06        & +12:09:48    & 51.293	& -8.404	   &  5.0 & $D$ & (c.2)\\
NGC 6858       &          & 20:02:56        & +11:15:30    & 51.360	& -10.304      & 10.0 & $D$ & (c.2)\\
NGC 6950       &          & 20:41:04        & +16:37:06    & 61.107	& -15.198      & 15.0 & $D$ & (c.2)\\
NGC 7084       &          & 21:32:33        & +17:30:30    & 69.963	& -24.302      & 16.0 & $D$ & (c.2)\\
\hline                                  
\end{tabular}
\end{table*}

\section{Methods}
\subsection{Mathematical model}
The traditional way of conducting membership assignment originates in the seminal works of \cite{Vasilevskis1958} and \cite{Sanders1971}. Those two works provided the basic ideas for developing parametric membership analysis. The derived mathematical model is based on the assumptions that there are two distinct kinematic populations in the observational field of the cluster and that the proper motion distributions of those two populations can be parametrised by bivariate Gaussians.

In these studies, an elliptical function was used to describe the field population's proper motion, while a circular one was used for the cluster. Nonetheless, if either an external gravitational influence or tidal effects were to act on different parts of the cluster, the cluster's proper motion distribution could be significantly affected, causing it to deviate from a circular function. For this reason, we tested four parametrisations of the probability distribution functions (PDF) in this study.

We adopted three variations of the general form of the PDF established by \cite{ZhaoHe1990} to take account of the observational errors in each individual point. The proper-motion dispersion parameters obtained from these PDFs are therefore the cluster and field intrinsic ones, which are independent of the observational errors in the proper motions. The first variation is a circular distribution, which is the most accurate representation of the PDF for a non-disturbed object with symmetric observation errors. The second variation is elliptical (allowing for the correlation coefficient) and the last a PDF in which we assume that the intrinsic dispersion cannot be observed because of the size of the errors \citep[as adopted by][]{BalaguerNunez2004}. We also tested a fourth PDF in which the $(\alpha, \delta)$ positions of the individual stars are taken into account, as in \cite{ZhaoChenWen06}. Nonetheless, unlike this study, which used the radial distance of the star to the centre of the cluster, we directly used the stars' coordinates and considered the cluster centre as a free parameter.

The adopted PDFs can therefore be written as a mixture of proper-motion ($\Phi$) and position ($\Psi$) PDFs

\begin{equation}
\Phi_t = \Phi_c\Psi_c + \Phi_f\Psi_f,
\end{equation}

\noindent
where $c$ and $f$ correspond to cluster and field parameters. 

The proper motion PDFs, $\Phi_c$ and $\Phi_f$, are assumed to be Gaussians of the form

\begin{eqnarray}
\label{eq:phiparam}
\Phi(\mu_{\alpha,i}, \mu_{\delta,i}) = \frac{1}{2\pi(1-\rho^2)^{1/2}(\sigma_{\mu_\alpha}^2+\epsilon_{\mu_{\alpha, i}}^2)^{1/2}(\sigma_{\mu_\delta}^2+\epsilon_{\mu_{\delta, i}}^2)^{1/2} } \nonumber\\
\exp{\Biggr\{-\frac{1}{2(1-\rho^2)} \Biggr[\frac{(\mu_{\alpha, i}-\mu_{\alpha})^2}{\sigma_{\mu_\alpha}^2+\epsilon_{\mu_{\alpha, i}}^2} + \frac{(\mu_{\delta, i}-\mu_{\delta})^2}{\sigma_{\mu_\delta}^2+\epsilon_{\mu_{\delta, i}}^2}} \nonumber \\
- \frac{2\rho(\mu_{\alpha, i}-\mu_{\alpha})(\mu_{\delta, i}-\mu_{\delta})}{(\sigma_{\mu_\alpha}^2+\epsilon_{\mu_{\alpha, i}}^2)^{1/2}(\sigma_{\mu_\delta}^2+\epsilon_{\mu_{\delta, i}}^2)^{1/2}}\Biggr] \Biggr\}.
\end{eqnarray}

\noindent While the $\Psi_c$ and $\Psi_f$ functions depend on whether we take into account the position of the stars or not. We refer to as 2D implementations those that do not take into account the positions, and 4D those that do. These parametrisations are given by

\begin{equation}
\label{eq:psiparam1}
\Psi_c = \left\{ \begin{array}{ll} 
n_c & \textrm{, 2D}\\
\frac{1}{1+g^{-1} \exp{\left\{-\frac{1}{2(1-\rho_{pos}^2)} \left[\frac{(\alpha_i-\alpha)^2}{\sigma_{\alpha}^2} + \frac{(\delta_i-\delta)^2}{\sigma_{\delta}^2} 
- \frac{2\rho_{pos}(\alpha_i-\alpha)(\delta_i-\delta)}{\sigma_{\alpha} \sigma_{\delta}}\right] \right\}}} & \textrm{, 4D}\\
\end{array} \right. 
\end{equation}

\begin{equation}
\label{eq:psiparam2}
\Psi_f = \left\{ \begin{array}{ll} 
\left(1-n_c\right) & \textrm{, 2D}\\
\frac{1}{1+g\exp^{-1}{\left\{-\frac{1}{2(1-\rho_{pos}^2)} \left[\frac{(\alpha_i-\alpha)^2}{\sigma_{\alpha}^2} + \frac{(\delta_i-\delta)^2}{\sigma_{\delta}^2} 
- \frac{2\rho_{pos}(\alpha_i-\alpha)(\delta_i-\delta)}{\sigma_{\alpha} \sigma_{\delta}}\right] \right\}}} & \textrm{, 4D}
\end{array} \right. 
\end{equation}

In Eq. \ref{eq:phiparam} above, $\mu_{\alpha, i}$, $\mu_{\delta, i}$ represents the proper motion of the $i-$th star, $\epsilon_{\mu_{\alpha, i}}$, $\epsilon_{\mu_{\delta, i}}$ the proper motion errors of the $i-th$ star, $\mu_{\alpha}$, $\mu_{\delta}$ the mean proper motions of the cluster and field stars (depending on the index of $\Phi$), $\sigma_{\mu_\alpha}$, $\sigma_{\mu_\delta}$ the intrinsic proper-motion dispersions of the cluster and the field, and $\rho$ represents the correlation coefficients of the cluster and field proper motion distributions. In Eqs. \ref{eq:psiparam1} and \ref{eq:psiparam2}, $n_c$, $1-n_c$ are the fraction of the cluster and field stars, $g$ is the ratio of the cluster to the field stars, $\alpha_i$, $\delta_i$ are the positions of the $i-$th star, $\alpha$, $\delta$ are the position of the cluster, $\sigma_{\alpha}$, $\sigma_{\delta}$ represent the dispersion, and $\rho_{pos}$ the correlation coefficient of the position distribution. 

The parametrisations presented above leave us with a vector $\mathbf \Theta$ of 11 or 16 parameters to be determined, depending on whether we consider 2D or 4D distributions.

\subsection{Model parameter estimation}
During the construction of the PM2000 catalogue, a global method was used to solve the star's astrometric parameters by selecting mean epochs such that the catalogue covariance matrices are diagonal, and the parameters obtained can be considered independent and identically distributed (iid). We are therefore allowed to write the likelihood function for a given region with $N$ stars as

\begin{equation}
\mathcal{L}=\prod^N_{i=1}\Phi_t\left(\alpha_i, \delta_i, \mu_{\alpha,i}, \mu_{\delta,i}, \mathbf\Theta\right).
\end{equation}

\noindent Following the maximum likelihood principle the most probable $\mathbf\Theta$ is that for which $\mathcal{L}$ takes its maximum value. A monotone transformation does not affect a maximum, and we can therefore write $\hat{\mathbf\Theta} = \arg\:\!_\mathbf{\Theta}\max \mathcal{L} = \arg\:\!_\mathbf{\Theta}\max \hat{\mathcal{L}}$, where
 
\begin{equation}
\hat{\mathcal{L}} = \sum_{i=1}^{N} \log \Phi_t\left(\alpha_i, \delta_i, \mu_{\alpha,i}, \mu_{\delta,i}, \mathbf\Theta\right)
\end{equation}

\noindent Usually the $\hat{\mathbf\Theta}$ is found to be the solution of a non-linear system of equations constructed from $\partial\hat{\mathcal{L}}/\partial\mathbf\Theta=0$, in which some sort of iterative procedure is used \citep[as in][]{Sanders1971, ZhaoHe1990, Dias06}. However, this approach depends on the initial chosen value of each unknown parameter. Therefore, the solution to the problem will converge to the real, physical one, provided that there are no local optima, or the initial value is not far from the physical optimum. We note that the last condition may be cluster dependent, and for many objects, we will indeed have local optima. Thus, one cannot guarantee that the obtained solution corresponds to the physical one.

To help us obtain global optima, we chose to use a strategy based on evolutionary computing to solve the optimization problem, the genetic algorithm (GA). This is an adaptive search heuristic for finding optimal solutions, which has been used several times in astronomy \citep{Hetem07, Howley08}. However, the number of generations necessary to attain convergence can be quite large, since GAs are generally slow to converge. On the other hand, these algorithms are usually very good at finding the region of global maximum. In this study, we decided to use an optimiser available in the R environment \citep{R2008} called RGENOUD, which is a mixture of a GA and a derivative-based hill-climbing algorithm. The GA identifies the region of the global maximum, while the quasi-Newton hill-climbing algorithm Broyden-Fletcher-Goldfarb-Shanno (or BFGS) based on the fitness function's derivatives, is used to optimise the best (higher likelihood) solutions and, therefore, find the region's maximum (which may be a global maximum). For a complete description, we refer to \cite{Mebane07}.

In principle we could allow the algorithm to search for a solution in $\mathbb{R}^{11}$ or $\mathbb{R}^{16}$ freely. Nonetheless, we know that, physically, the search region can be constrained for some of the parameters, such that if an optimal point were to exist outside this region, it would not have any physical meaning. For example, if most cluster stars were inside the extracted region, we could assume that the position of the cluster center would never be smaller or greater in value than the positions of the most extreme stars in that region. The minimum and maximum of the cluster and field proper-motion components should never be smaller or greater than the minimum and maximum proper-motion components of the stars themselves. We should also ensure that the correlation coefficient moduli never equals unity, otherwise we could encounter problems related to divisions by zero. We also constrain the cluster's proper motion dispersions to have a maximum of 5 mas yr$^{-1}$, which is an estimated upper limit to the average errors in the individual proper motions. Finally, we allow the field's proper motion (and cluster's position) dispersions to be almost free, adopting a very high limit of six times the standard deviation of the individual proper motions (and stellar positions).

Hence, given the four $\mathbb{R}^1$ sets $\mathbf{M_{\alpha, \delta, \mu_\alpha, \mu_\delta}}$ of all the stellar positions and proper motions in the region, we adopted the following constraining conditions for the unknown parameters:

\begin{equation}
\hat{\mathbf\Theta}
\mathrm{\;is\; such\; that\; }
\left\{
\begin{array}{lc}
\min\left(\mathbf{M_{\mu_\alpha}}\right) \le \mu_{\alpha, c/f} \le \max\left(\mathbf{M_{\mu_\alpha}}\right) &\\
\min\left(\mathbf{M_{\mu_\delta}}\right) \le \mu_{\delta, c/f} \le \max\left(\mathbf{M_{\mu_\delta}}\right) &\\
0.01 \le \sigma_{\mu_{\alpha, c}} \le 5.00 &\\
0.01 \le \sigma_{\mu_{\alpha, f}} \le 6\cdot\sigma\left(\mathbf{M_{\mu_\alpha}}\right) &\\
0.01 \le \sigma_{\mu_{\delta, c}} \le 5.00 &\\
0.01 \le \sigma_{\mu_{\delta, f}} \le 6\cdot\sigma\left(\mathbf{M_{\mu_\delta}}\right) &\\
-0.99 \le \rho_{c/f} \le 0.99 &\\
&\\
0 \le n_c \le 1 & \mathrm{,\;if\;2D}\\
0.01 \le g \le \mathrm{length}\left(\mathbf{M_{\alpha}}\right) & \mathrm{,\;if\;4D}\\
\min\left(\mathbf{M_\alpha}\right) \le \alpha \le \max\left(\mathbf{M_\alpha}\right) &\mathrm{"}\\
\min\left(\mathbf{M_\delta}\right) \le \delta \le \max\left(\mathbf{M_\delta}\right) & \mathrm{"}\\
0.01 \le \sigma_\alpha \le 6\cdot\sigma\left(\mathbf{M_\alpha}\right) &\mathrm{"}\\
0.01 \le \sigma_\delta \le 6\cdot\sigma\left(\mathbf{M_\delta}\right) &\mathrm{"}\\
-0.99 \le \rho_{pos} \le 0.99 &\mathrm{"}\\
\end{array}
\right.
\end{equation}

\noindent
where $\sigma\left(\mathbf{M_{\alpha, \delta, \mu_\alpha, \mu_\delta}}\right)$ represents the standard deviation of the respective set of parameters.

When a solution is found, the Hessian matrix ($\mathcal{H}$) is computed for the solution parameters. We use this matrix to estimate the individual parameter's errors in the optimization procedure; this is necessary because for a regular problem \citep[in the sense of][]{Wald49}, as considered here, when one maximises the log-likelihood function, the covariance matrix on the parameters is closely approximated by the negative of the inverse of the Hessian matrix at the solution point \citep[see][]{Pawitan01}. Thus, the vector of the variances in 
the parameters is estimated to be

\begin{equation}
\hat{\mathbf\Theta}_{var}\approx\mathrm{diag}\left(-\mathcal{H}^{-1}\right),
\end{equation}

\noindent
and we use the square root of the elements of this vector to estimate the fitting procedure errors.

Once the cluster and field PDF's unknowns are found, the membership problem is solved because we are then able to compute $p_i$, the cluster membership probability of the $i$-th star

\begin{equation}
\label{eq:prob}
p_{i(\mu_{\alpha,i}, \mu_{\delta,i}, \mathbf\Theta)} = \frac{\Psi_c\Phi_c}{\Psi_c\Phi_c + \Psi_f\Phi_f}.
\end{equation}

\noindent The number of apparent cluster members can be estimated from the mixture proportion $n_c$ in the following way. First, from the list of stars used during the optimization process (the list with the quality cuts explained in Sect. 2.2), a sublist of the $N_{opt}*n_c$ most probable stars (where $N_{opt}$) is created. From this list of most probable stars, the probability of the least probable star $P_{min}$ is obtained. Finally, the number of members in the extracted region (without any probability cut applied) can be estimated from the number of stars in that region with membership probabilities greater than $P_{min}$.

\section{Validation}

We verify the reliability of our work in four steps. We firstly validate the automatic optimization method using very precise data. Secondly, we test the new 4D parametrisation by comparing it with another similar parametrisation and method used before in the literature. Thirdly, we validate the data by comparing our results for a well known cluster with the literature. Finally, we perform a comparison of individual members for one ``A" and one ``B" class cluster.

\subsection{Optimisation method}

Very precise proper motion data are available for the open cluster NGC 1817. These data were presented by \cite{BalaguerNunez2004} (hereafter BN04), in which proper motions were determined from 25 plates covering a total time-span of 81 years. The mean error in the data of more than 80\% of the stars is $\epsilon_\mu=1.55$ mas yr$^{-1}$, while it is $0.97$ mas yr$^{-1}$  for 32\% of the stars, for objects as faint as $V<16$.

Since we use a new optimization method in this work, we tested it using these proper motion data. Since we consider the method itself, we used the same parametrisation as in BN04, and set the intrinsic dispersion of the cluster Gaussian PDF to zero. We also applied the same cuts in the data ($|\mu|<30$ mas yr$^{-1}$). As can be noticed in Table \ref{table:NGC1817modval}, the agreement between our results is quite remarkable, indicating that the method adopted herein allows us to correctly determine the parameters in a fully automatic manner.

We also performed extensive tests using Monte Carlo randomly generated sets of Gaussian mixtures. In virtually all cases, the parameters of the distributions were recovered correctly.

\begin{table}
\caption{Method validation for BN04's NGC 1817 data (units of $\mu$ and $\sigma$ are mas yr$^{-1}$). }             
\label{table:NGC1817modval}
\centering
\begin{tabular}{l c c c c c c}
\hline\hline
Work& $n_c$ &$\mu_{\alpha}\cos\delta$ & $\mu_{\delta}$ & $\sigma_{\mu_{\alpha}\cos\delta}$ & $\sigma_{\mu_{\delta}}$ & $\rho$ \\
\hline
Our	     & $0.252$   	&$0.25$& $-0.94$ 	     & & & \\ 
method & $\pm0.02$&$\pm0.11$& $\pm0.07$ & & & \\
		& 	           		& $2.33$& $-4.08$   & $5.11$       & $5.93$       & $-0.04$ \\ 
		& 		  		& $\pm0.24$& $\pm0.27$ & $\pm0.18$ & $\pm0.20$ &$\pm0.04$ \\      
&&&&&&\\
BN04  & $0.261$   	& $0.29$& $-0.96$      	& & & \\ 
		&$\pm0.02$	& $\pm0.10$& $\pm0.07$ & & & \\ 
		& 			  	& $2.29$& $-4.25$      & $5.69$       & $6.38$       & $-0.08$ \\ 
		& 		  		& $\pm0.02$& $\pm0.27$ & $\pm0.02$ & $\pm0.14$ &$\pm0.03$ \\
\hline
\end{tabular}
\end{table}

\subsection{4D parametrisation}

To test the 4D parametrisation, we compared our results for a set of open clusters with those obtained using another parametrisation and method. The Stochastic Expectation Maximization (SEM) method \citep{SEM} can be used to deblend the two components corresponding to the field and cluster, assuming that the data can be described by a mixture of 4D Gaussians. The SEM algorithm is non-informative and 
iteratively solves the  maximum  likelihood  equations,  with  a stochastic step, for a multivariate mixture of  Gaussian distributions. It has been adopted to deconvolve the thin and thick disk populations from velocity distributions \citep{Soubiran1993, Soubiran2003}. The algorithm initiates at a position decided at random, assuming that the sample is a mixture of 2 discrete components. We repeat the process 150 times. Most of the time, the algorithm converges to the same solution. When multiple solutions are found, the most frequent one is adopted.

\begin{figure}
   \centering
   \includegraphics[width=9cm]{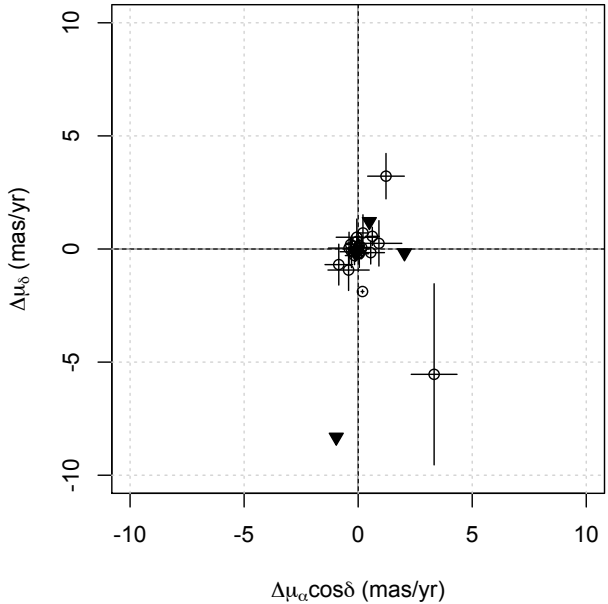}
      \caption{Comparison between the result of the fit using the method described in this work and SEM. The three filled triangles indicate points with error bars greater than the plot axis.}
         \label{fig:gz_vs_sem}
   \end{figure}

We used SEM to analyse all the regions on our list. Afterward, we compared the solutions with those obtained by the 4D parametrisation described herein. The average compatibility between the results, presented in Fig. \ref{fig:gz_vs_sem} is very good, with a clear distribution centred on $(0.25\pm0.9; -0.46\pm2.17)$ mas $\textrm{yr}^{-1}$ before a simple $3\sigma$ rejection to eliminate outliers, and centred on $(0.12\pm0.64; -0.13\pm1.45)$ mas yr$^{-1}$ after the rejection. 

\subsection{Data}

To test the data used throughout this work, we applied the method described above to the 4D parametrisation to obtain the kinematic parameters for the open cluster NGC 2682. This object, also called M67, is an old, $\sim$4 Gyr and well known open cluster that has been extensively studied in the literature \citep[][and references therein]{Yadav2008}. 

By comparing the results obtained for this open cluster with SEM and the new method described above, we find that they are in close agreement, generally less than the fitting errors computed for those parameters: $\Delta(\mu_\alpha\cos\delta, \mu_\delta) = (0.06, 0.02)$ mas yr$^{-1}$, $\Delta(\sigma_{\mu_\alpha\cos\delta}, \sigma_{\mu_\delta}) = (0.20, 0.10)$ mas yr$^{-1}$, and $\Delta\rho = 0.02$.

The comparison of the kinematic parameters derived from our method with those from previous studies in the literature is shown in Table \ref{table:NGC2682Table}. We observe very good agreement between all the results within the estimated errors, although we note that some of the errors in that table are estimated from the standard deviations of member stars, and not from the fitting procedures. 

\subsection{Individual membership}

We verify that the members determined by our method are compatible with those found in previous studies by comparing the members obtained for our ``A" class cluster (NGC 2682) and one ``B" class cluster (NGC 7036) with previously published membership lists. In performing this analysis, we adopted the parametrisation that does not take into account the cluster's internal dispersion.

For NGC 2682, \cite{Yadav2008} (hereafter Y08) computed the membership probabilities of 2410 stars in its vicinity. Their study was relatively deep in magnitude, to V$\sim$20, although probably does not probe the entire cluster because of its large angular extension. Unfortunately, the authors did not indicate the number of members found, but a cut at 60\% in membership probability was adopted in their work. For this value, 595 stars would be considered as members. When analysing PM2000 data, we obtained 502 members from a total of 1386 stars. We note that our catalog is much brighter than that of Y08, but covers a more extended area.

Between those two membership lists, there are 271 stars in common. Nonetheless, to perform comparison, we need to take into account that only 553 stars are common between our initial catalogues, and that among these common objects, 313 are considered to be members by ourselves, while 402 are Y08 members. This means that $\sim$87\% of our members that could be listed as members in Y08 were listed as such, while only $\sim$68\% of Y08 members were listed by ourselves. 

Since PM2000 proper motions are precise in this particular field, with $\sigma_\mu\le2$ mas yr$^{-1}$ for about 50\% of them, but in the same magnitude range (V$\le$16) Y08 has only about 28\% of its stars with such small errors, we attribute this lower value of $\sim$68\% to a possible contamination of field stars among Y08 members.

For NGC 7036, \cite{Dias06} (hereafter D06) analysed 69 stars in its vicinity, obtaining 20 members for this possible open cluster remnant. We analysed 91 objects close to NGC 7036, also obtaining 20 members. Adopting the same procedure as applied to NGC 2682, we found that $\sim$65\% of our members that could be listed as members in D06 were listed as such, while $\sim$68\% of D06 members were among our members. 

Taking the results for these two clusters in to consideration, we conclude that there is relatively good agreement between the members obtained by our method and data with those previously published.

\begin{table}
\caption{Kinematic parameters determined for NGC 2682.}
\label{table:NGC2682Table}
\centering                          
\begin{tabular}{l c c} 
\hline\hline               
Reference & $\mu_{\alpha}\cos\delta$ & $\mu_{\delta}$ \\
& (mas yr$^{-1}$) &(mas yr$^{-1}$) \\
\hline                     
	This work 					& $-8.32\pm0.07$ & $-5.65\pm0.07$ \\
	\cite{Frinchaboy2008} 	& $-7.87\pm0.61$ & $-5.60\pm0.59$ \\
	\cite{Dias07} 				& $-8.62\pm0.28$ & $-6.00\pm0.28$ \\
	\cite{Kharchenko2005} 	& $-8.31\pm0.26$ & $-4.81\pm0.22$ \\
\hline                                  
\end{tabular}
\end{table}

\section{Results}

We applied the optimization method and all variations in the parametrisation described in Sect. 3.1 to the extracted PM2000 data. We thus obtained four solutions for each cluster, and have to determine the most reliable. We first verify whether each solution is in agreement with all others. Next, we visually inspected  probability histograms, the vector point diagrams (VPD) and star charts of members and non-members stars, and the DSS images of all clusters, and classified the results as ``good'', ``intermediate", or ``poor''. To avoid any prejudice, we performed a blind classification: the names of the objects were not written on the diagrams and images while they were being analysed. During this check, we noticed that the probability histogram exhibited the expected field-cluster distribution with a dual peak distribution in most cases. Furthermore, in all cases, the VPD diagrams were indicative of accurate fits. Hence, we decided not to use those two diagrams when assessing the physical reliabilities of the four solutions

Although this classification of the solutions was subjective, we adopted two clear criteria: some kind of structure or central concentration of stars in the member star chart (in contrast to an almost homogeneous distribution for non-members) and some correlation between a certain magnitude range and the member stars. The solutions were classified as ``good" when a clear concentration of member stars was present (even if there was some contamination), and as ``intermediate" when the members were not so clear concentrated (or when the concentration was very off-centred), but there was some correlation between the members and a certain magnitude range. Finally, they were classified as ``poor" when the members were sparsely distributed without any apparent correlation with any magnitude range.

For NGC 2682, a reference open cluster for which we obtained ``good'' results,  it is clear from the member diagrams that the vast majority of its stars were correctly classified, even if the inspection of its non-member diagrams indicates that some of this cluster's members were probably not classified as such. Moreover, we were able to obtain good results for the 4D parametrisation in this cluster, because it is a well-populated cluster in the magnitude range of PM2000, which allowed a correct determination of its centre and physical dispersion, and its kinematic parameters.

However, of the four parametrisations that we tested, the PDF which considers the internal dispersion of the clusters to be too small to be observed, provided the least number of results classified as ``poor''. This can be explained physically by the data of the distant clusters being affected by this dispersion in the proper motion far less significantly than the errors in the individual measurements, as noted by \cite{BalaguerNunez2004}.

After \cite{Gunnetal1988}, the theoretical prediction for the intrinsic velocity dispersion for a Hyades like open cluster in a state of dynamical equilibrium is 0.23 km s$^{-1}$, and the observed value based on Hipparcos data for the Hyades cluster is 0.3 km s$^{-1}$ \citep[][and references therein]{deBruijne2001}. \cite{Mamajek2010} reported an intrinsic dispersion of 1 km/s for several nearby objects, and an upper limit of 1.1 km s$^{-1}$ in $\alpha$ Persei was set by \cite{Makarov2006}. Using the latest proper motion catalogues, it would be unrealistic to probe values similar to those for any relaxed open cluster that is neither located at very nearby distances nor is highly populated: at only a few hundred parsecs, the dispersion is already at $\mu $as yr$^{-1}$ scales. Thus, the use of this additional degree of freedom in the analysis could allow the likelihood function to reach a better fitness value when modelling well the field distribution than when segregating field from cluster stars.

\addtocounter{table}{1}

We present the results obtained by the PDF that considers the internal dispersion of the clusters to be too small to be observed, for our entire input list in Table \ref{table:pdfparameters}. It includes all the fitted parameters from the cluster and the field PDFs, as well as their associated errors. In this table, we also present an estimate of the number of apparent members, and the reference class from Table 1. This table is divided into three sections, corresponding to the quality of the solutions (``good", ``intermediate", ``poor").

\addtocounter{table}{1}

In Table \ref{table:example} (fully available only through CDS), we present the membership probability information (computed from Eq. \ref{eq:prob}) for all the stars in the extracted regions. In addition, we include some columns from the PM2000 catalogue, as well as the object's 2MASS identifier when available.

\begin{figure}
   \centering
   \includegraphics[width=9cm]{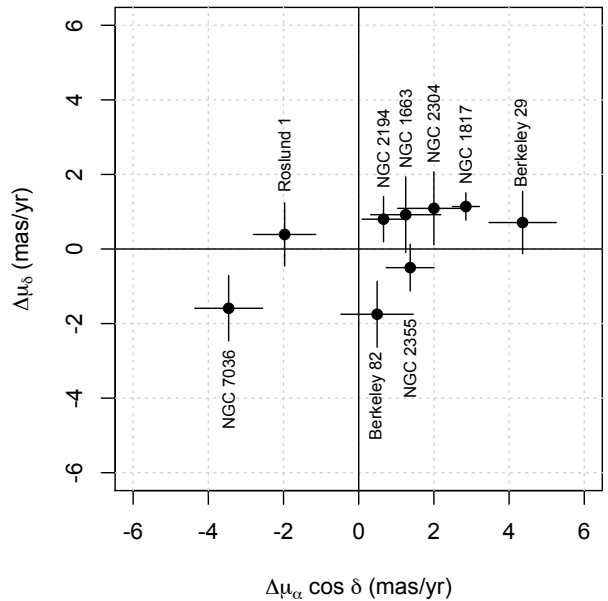}
      \caption{Comparison between some results of this work and those published in the D06 catalogue. The differences in the open cluster proper motions are in the sense of hereof minus D06 results.}
         \label{fig:gz_vs_dias06}
   \end{figure}

For some clusters on our list, D06 also obtained kinematic parameters by using UCAC2 data \citep{Zacharias2004}. For those objects, we compare our proper motions with those of D06 work in Fig. \ref{fig:gz_vs_dias06}. We note that the objects tend to have systematically positive values of $\Delta\mu_\alpha\cos\delta$ and $\Delta\mu_\delta$ alike, which means that their proper motions determined by PM2000 data are somewhat higher than those determined by UCAC2 data, by $(0.8, 0.1)$ mas yr$^{-1}$. The same was found for all the other three parametrisations.

\begin{figure}
   \centering
   \includegraphics[width=9cm]{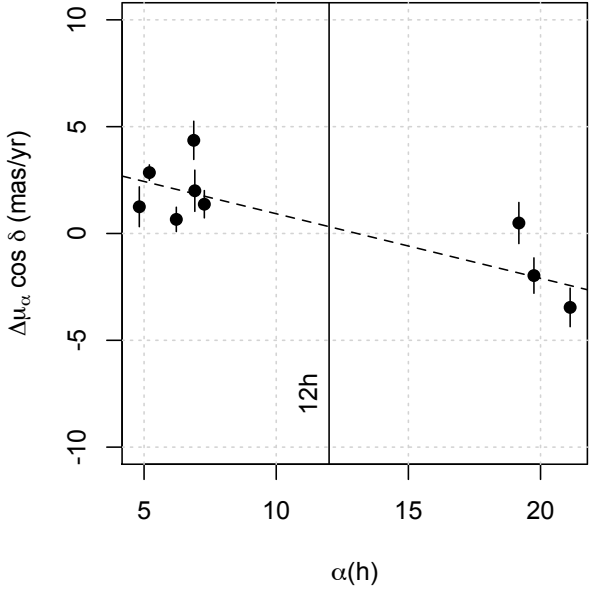}
      \caption{Systematic effect in the results for the common clusters between this work and the D06 catalogue. The dashed line is a weighted least-squares fit on the points.}
         \label{fig:cluster_systematics_weighted}
   \end{figure}

These results were surprising to us, so we performed several tests using the comparison data. During those tests, we found that the proper motion differences $\Delta\mu_\alpha \cos\delta$ were heavily correlated with the cluster's right ascension. This systematic effect can be clearly seen in Fig. \ref{fig:cluster_systematics_weighted}, in which we present the differences between the results of the parametrisation that does not take into account the cluster's internal dispersion and the D06 catalogue.

\begin{figure}
   \centering
   \includegraphics[width=9cm]{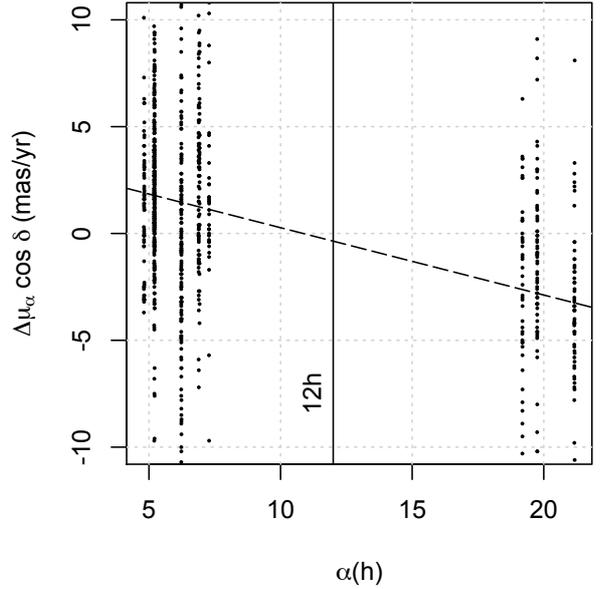}
      \caption{Differences in proper motions between common PM2000 and UCAC2 stars in the sense PM2000 minus UCAC2. This comparison includes only stars in the regions of the common open clusters reduced by this work and D06. The dashed line is a least squares fit to the points.}
         \label{fig:stars_pm2000_minus_ucac2}
   \end{figure}

We concluded that the origin of this systematic effect was the data itself, since the $(\alpha, \delta)$ position of the object was not used in 3/4 of our parametrisations, it was not used in D06 reduction, and the results of all the four parametrisations exhibited exactly the same trends. We therefore directly compared the proper motions of the common stars between PM2000 and UCAC2 (the proper motion catalogue used by D06) inside the cluster regions, and noted a clear dependence on the right ascension, which can be seen in Fig. \ref{fig:stars_pm2000_minus_ucac2}. 

\begin{figure*}
   \centering
   \includegraphics[width=\linewidth]{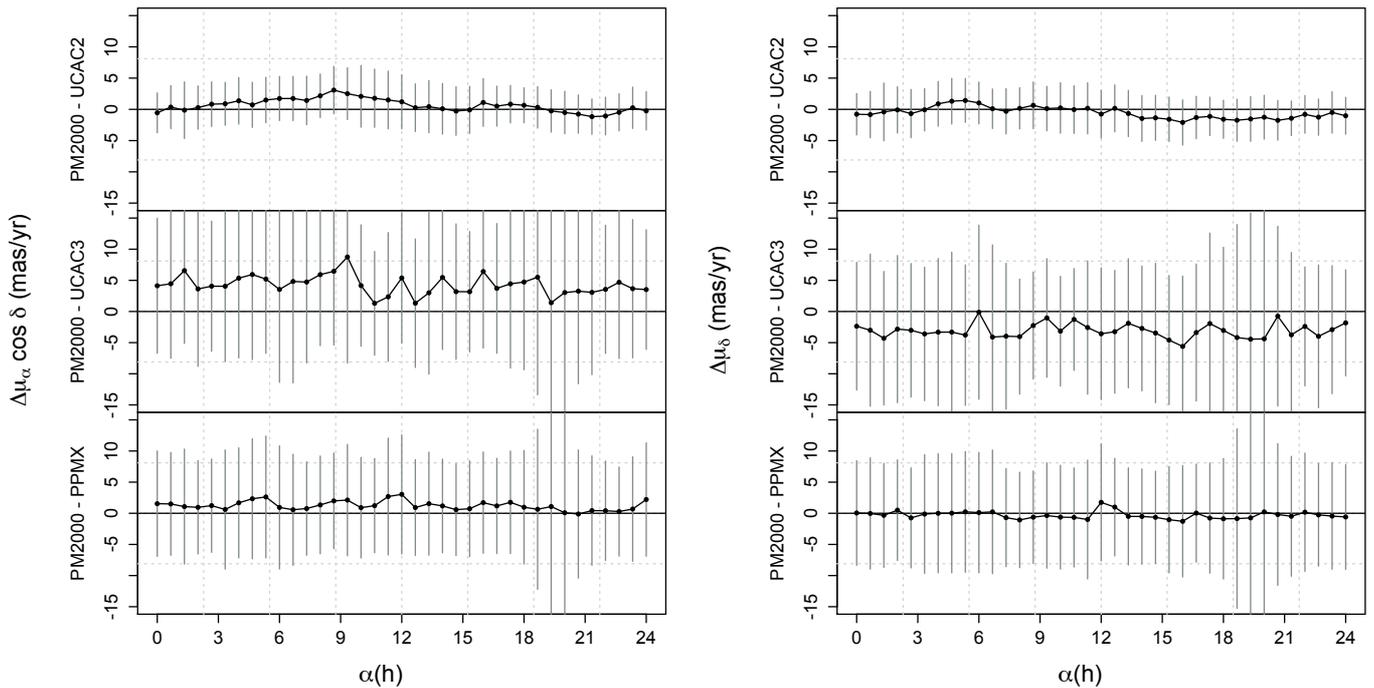}
      \caption{Average differences in bins of $\alpha$ in the proper motion components of common stars between PM2000 and UCAC2, UCAC3, and PPMX catalogues.}
         \label{fig:av_pm2000_minOthers}
   \end{figure*}

We then compared the entire PM2000, UCAC2, UCAC3 \citep{Zacharias09}, and PPMX \citep{Roser08} catalogues (Fig. \ref{fig:av_pm2000_minOthers}). For UCAC2, there appears to be a periodic systematic variation. However, when comparing each of them with Tycho-2, no similar effect was detected, indicating that this problem affects mostly faint stars, whose first epoch data for deriving proper motions were neither AC2000 nor Cartes du Ciel plates. 

For UCAC3, we measure strong offsets in both proper motion components ($4.3\pm1.6$ mas yr $^{-1}$ in $\alpha$ and $-3.1\pm1.2$ mas yr$^{-1}$  in $\delta$), while relative to PPMX there is a small offset in the right ascension component ($1.2\pm0.75$ mas yr$^{-1}$), but no noticeable effect in declination. Since all these four catalogues use common material and cannot be considered to be completely independent, no firm conclusions could be drawn from a simple analysis and a detailed study is required to address the origin of these effects.

For clusters classified as ``non-existent'', ``doubtful'' or ``not found'' in the D07 catalogue, the chosen parametrisation failed to provide a good fit (our visual classification of the solutions was ``poor''). This may indicate that those regions probably are not composed by two different kinematic components, thereby confirming the flag ``non-existent" of the D07 catalogue. This is important, since the aforementioned flag is based mainly on a visual inspection performed by \cite{Sulentic1973}. These objects are the NGCs 2224, 2234, 2265, 6525, 6738\footnote{Confirming the results of \cite{Boeche2003}, which was based on Tycho-2 data.}, 6837, 6839, 6840, 6843, 6858, 6950, and 7084. We note that this ``poor'' classification cannot be caused by the limiting magnitude of our catalogue since NGC objects should be relatively bright. The cluster Dol Dzim 2 and Dolidze 35, with the flags ``not found'' and ``doubtful'' in the D07 catalogue were also classified as presenting a ``poor'' solution.

Regarding all the other clusters, we obtained solutions classified as ``good" for the NGCs 1817, 2169, 2194, 2355, 2682, and Berkeley 82. The clusters for which we obtained solutions considered as ``intermediate'', and therefore require further analysis, likely by using additional information from photometric, radial velocity data and deeper or more precise proper motions, are: Berkeley 43, Berkeley 45, Berkeley 47, the Chupina series, King 26, the NGCs 1807, 2678, 7036, Roslund 1, Skiff J0614+129, and Teutsch 11.

In the case of NGC 2678, we note that the stars classified as members are concentrated in direction of NGC 2682. The brightest stars in this region (those that visually define the cluster) were also not classified as members. Since these two clusters are close to each other, this may indicate that these stars are members of NGC 2682 and that NGC 2678 does not exist.

The remaining objects on our list were classified as ``poor". In the case of Berkeley 29, even though we have around 80\% of the stars with membership probability greater than 51\% in common between our and D06 analyses, the cluster's proper motions in right ascension are not compatible with those published in D06. We notice that both solutions are probably only good fittings of the field distribution, since from the photometric work of \cite{Tosi2004} we can see that the vast majority of this open cluster's stars are at $V>18$, far beyond the reach of the PM2000 (used here) or UCAC2 (used in D06) catalogues. For NGC 2395, we could not distinguish a concentration of member stars, and the bright stars that define the cluster were not considered as members.

For all the other solutions classified as ``poor'', we examine with regions in which the difference between the number of field and the cluster stars was very high. As a result, the natural tendency of the likelihood function was to reach a better fitness value for the determination of a field only distribution. Nonetheless, we note that some clusters classified as ``poor'' can have a good determination of their proper motions, as in the case of NGC 7772, which we comment on the next section.
     
\subsection{Comments on selected objects}

Because of the physical interest of some open clusters in this study, we briefly discuss the properties of a small subset of them individually.

\subsubsection{NGC 1807}
The NGC 1807 open cluster is a concentration of stars close to NGC 1817. The solutions obtained for this object using all the four parametrisations indicate that its most probable members are concentrated in one border of the extracted field, mainly in the direction of NGC 1817. For the 4D parametrisation, the concentration in that part of the star chart is even larger than the other ones. In addition, the fitting parameters obtained for the cluster's PDF are compatible (within the fitting errors) with those obtained for NGC 1817 (as can be seen in Table \ref{table:pdfparameters}). 

We conclude that the existence of a separate cluster from NGC 1817 in the NGC 1807 region is not supported by our analysis of PM2000 kinematical data. This upholds \cite{BN04b, BalaguerNunez2004} assertion, that there is actually only one extended cluster in that region. However, we note that a radial velocity study would be of great value to definitely settle this issue.

\subsubsection{NGC 2194}
NGC 2194 is an open cluster located in a moderately rich field projected along the Galactic anti-centre direction, which has been studied by \cite{Kyeong2005}, \cite{Piatti2003}, and \cite{Sanner2000}. Only the final of these three aforementioned studies deals with proper motions. \cite{Sanner2000} were the first to analyse this object's proper motion. In their study, proper motions were obtained by using photographic plates from the Bonn Doppelrefraktor (first plate from 14.02.1917) and from CCD observations performed at the 1.23m telescope at the Calar Alto Observatory (15.10.1998). They determined the cluster's proper motion to be $(\mu_\alpha\cos\delta;$ $\mu_\delta) = (-2.3\pm1.6;$ $0.2\pm1.5)$ mas yr$^{-1}$ (the indicated errors are the fitted widths of the PDF).

This cluster was later analysed, automatically, in D06, in which its proper motion was computed from UCAC2 data ($-0.59\pm0.53;$ $-3.49\pm0.53$) mas yr$^{-1}$. As one can promptly see, there is a wide discrepancy between the proper motion values obtained by those two studies, even if compatible to within $3\sigma$. It is even more confusing that the entry in the D07 catalogue, ($-0.31\pm0.64;$ $-4.40\pm0.64$) mas yr$^{-1}$, is once again different\footnote{The same value is quoted in the WEBDA database, and was computed from Tycho-2 data in \cite{Dias02}}, even though it is compatible with the one obtained using UCAC2 proper motions.

We found herein the mean proper motion of this object to be more compatible with that of D06, with a value of ($0.07\pm0.20;$ $-2.69\pm0.29$) mas yr$^{-1}$. The almost $3\sigma$ discrepancy between the cluster's proper motion results obtained by this work, D06, and \cite{Sanner2000}, may be caused by the different materializations of the HIPPARCOS reference system. As for UCAC2, PM2000 is linked to HIPPARCOS as materialized by Tycho-2. In constrast, \cite{Sanner2000} used ACT stars to perform the transformation from plate to celestial coordinates, ACT being linked to HIPPARCOS, but as materialized by Tycho-1.

It is also interesting that the proper motion of this cluster is compatible with that obtained for Skiff J0614+129. We can also identify some of the brightest stars in the region of Skiff J0614+129 (those that define the central concentration in its member star chart) among the list of NGC 2194 members, but a more detailed analysis using photometric and radial velocity data would be required to check whether they are somehow connected.

\subsubsection{NGC 7772}
NGC 7772 was firstly proposed to be an open cluster remnant by \cite{Bica2001}, and confirmed to be one by \cite{Carraro2002}. In the latter study, NGC 7772's age was estimated to be 1.5 Gyr, and its distance from the Sun to be 1.5 kpc, which assigned the membership to 14 stars. Nonetheless, since that work was purely based on photometric data, Carraro himself warned that his analysis needed to be constrained by radial velocity and proper motion determinations. Here we constrain the membership of the stars in this object by using proper motions.

We note that the fitting of two kinematic populations to this cluster VPD, using the PDF that does not take into account the internal dispersion, does not allow a highly reliable segregation between the remnant and the field, since only two stars were classified as members using the criteria for the number of members presented at the beginning of this section ($n=2.34\pm1.56$). We believe that this to be caused by our choosing the PDF that does not take into account the internal dispersion between the member stars: since this is a remnant, the dispersion could be greater than the individual measurement errors, and we should be able to resolve it.

\begin{figure}
   \centering
   \includegraphics[width=9cm]{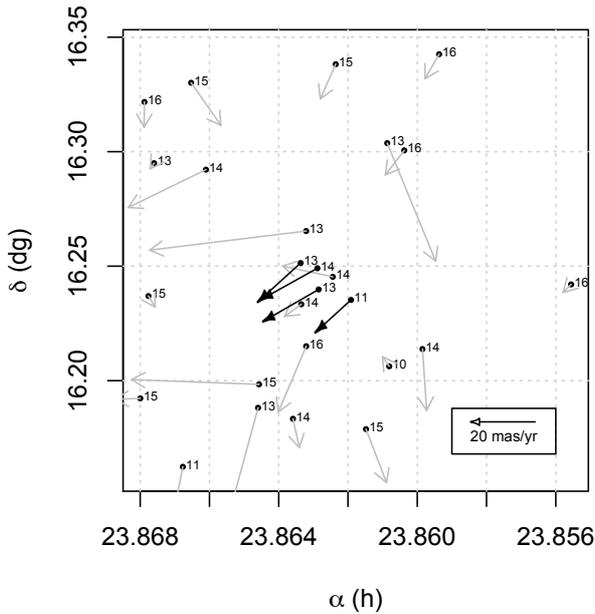}
      \caption{Star chart of the field around NGC 7772. The non-member stars have their proper motions represented by the grey vectors, and the members by black ones. Meridian $V$ magnitudes are also shown.}
         \label{fig:ngc7772_propermotion}
   \end{figure}

Since this is a sparsely populated region, we can, however, analyse a plot of the proper motions represented as vectors superimposed directly on the Star Chart, as can be seen in Fig. \ref{fig:ngc7772_propermotion}. This graphical representation allows us to observe what is happening from the point of view of the kinematic, spatial, and magnitude distributions. We realise that there exists a tightly packed spatial concentration of similar magnitude stars that share almost the same proper motion vector in this region.

On the basis primarily of their proper motions, we visually chose the four (compatible with $n$ to within $\sim1\sigma$) most similar stars in the high-concentration region, two of which are the member stars classified by our automatic reduction. We note that the other stars have higher membership probabilities in the 4D solution.

Since these are the constituent stars of this object, and whose errors in their individual proper motions are very small ($\langle\mu/e_\mu\rangle = 4\%$), the average proper motion is a very good estimate of this remnant's proper motion. The final average proper motion of the object is therefore $\langle\mu_{\alpha}\cos\delta\rangle = 14.90 \pm 3.13$, $\langle\mu_\delta\rangle = -10.65\pm1.11$ mas yr$^{-1}$, which is highly compatible with that determined automatically (see Table \ref{table:pdfparameters}). This is the first ever determination of a proper motion for this object.

\subsubsection{Berkeley 82}
Berkeley 82 is a sparsely populated open cluster that has yet to be thoroughly studied. This cluster is projected in the direction of a HI supershell \citep{KK2000} that has a kinematically derived distance of 1.4 kpc and a projected surface of 340 x 540 pc$^2$ (in the l, b directions). These authors suggested that this $\sim5$ Myr supershell may be physically related to some of the open clusters in its region: NGC 6738, Berkeley 43, and Berkeley 82. The existence of the first of these three objects was ruled out. Berkeley 43 perhaps is an interesting object without proper motion determination in the literature, but unfortunately our automatic solution for it was classified as ``intermediate" probably because of the faintness of this cluster.

However, for Berkeley 82, we were able to assign membership of stars to this object and compute its proper motion, which was measured to be ($3.02\pm0.61;$ $-1.64\pm0.48$) mas yr$^{-1}$. From D07, a Tycho-2 determination for the proper motion of this object was obtained by \cite{Loktin2003} during a study of the rotation rate of the Galaxy ($-0.06\pm0.99;$ $-4.38\pm1.78$) mas yr$^{-1}$. Nonetheless, no list has been published with individual membership probabilities, so one cannot explore the connection between the cluster and the shell on a star by star basis. According to WEBDA Berkeley 82 is located at a distance of 870 pc, indicating that this cluster could well lie at one of the borders of the shell (if the cloud size is taken into consideration).

Based on the quoted distance, our proper motion determination amounts to a tangential velocity of $\sim13.8\pm2.4$  km s$^{-1}$. \cite{KK2000}, determine the expansion velocity of the shell to be $\sim15$ km s$^{-1}$, and its central velocity to be $18$ km s$^{-1}$. The similarities between the velocities that they obtained for the HI supershell and that which we obtained for the open cluster are quite remarkable. A radial velocity study would undoubtedly be of great value to determine the space velocity of this object, and finally reach a conclusion about the physical connection between this cluster and the supershell.

\subsubsection{NGC 1663}
During their photometric analysis, \cite{Baume2003} suggested that NGC 1663 is a $\sim$2 Gyr possible open cluster remnant. In that study, the authors derived a list of 2 members and 4 possible members, from a classification based on UBVI photometric data. They noted, however, that it was difficult to determine the true nature of this object.

Our automatic solution for NGC 1663 was classified as ``poor", since we could observe neither a clear concentration of member stars, nor any correlation with the brightest stars, in the member star chart. Moreover, the member list obtained by our solution included only one star from the photometric member list.

Unfortunately, it is unclear whether members and possible members in \cite{Baume2003} show common proper motions. Because of the lack of ancient epoch data in this particular zone of PM2000, the proper motion errors are greater than the average of the catalogue. This lack of precision prevents us from drawing firm conclusions, and a more precise proper-motion study and/or a radial velocity analysis is needed to clarify the nature of this possible open cluster remnant.

\section{Conclusions}
We have developed a fully automatic system to determine the kinematic parameters and membership probability lists of open clusters using kinematic and spatial data. The adopted method permits us to take into account any additional parameter (such as radial velocity), given that an analytical form of its probability distribution functions for the cluster and field populations are known {\it a priori}. Using this tool, we visually compared the results for several possible parametrisations for the PDFs for all open clusters in PM2000. We concluded that the most reliable function is one that does not take into account the intrinsic dispersion of the cluster.
 
Based on PM2000 data, we obtained proper motions and kinematic membership lists for open clusters in the Bordeaux's PM2000 zone. For five of them, it was the first such measurement in the literature. We note, however, that for some of those objects, additional studies are necessary, if possible, based on radial velocity data to confirm the membership determination. Moreover, we confirmed from the kinematic point of view, the non-existence of 13 NGC objects, as well as Dol Dzim 2 and Dolidze 35, which are flaged as ``not found" and ``doubtful" in the D07 catalogue.
 
We concluded that the open cluster NGC 1807 probably does not exist, as previously suggested in the literature \citep{BN04b, BalaguerNunez2004}. Nonetheless, we note that a radial velocity study to finally settle this issue is still required. 
  
We also determined the tangential velocity of Berkeley 82 to be remarkably similar to the velocities associated with the HI supershell located near it. Thus a physical connection between both objects remains plausible.
  
By comparing PM2000 and UCAC2 proper motions, we found a periodic systematic variation as a function of the object's right ascension. Strong offsets in both proper motions components were also detected on comparing PM2000 with UCAC3. However, a comparison of PM2000 with PPMX showed only a small offset in the $\mu_\alpha\cos\delta$ component, and no noticeable effect in the $\mu_\delta$ component.

We finally conclude that the blind use of parametric fully automatic methods in present large-scale proper-motion catalogues is unreliable for most open clusters, since incorrect conclusions are often derived by inspecting VPD and probability histogram plots alone. When conducting studies based on these types of analyses, the results should be critically assessed using independent data; one should avoid using the aforementioned diagrams, since they are directly connected to the fitting itself. As a result, one only checks how mathematically successful the fit is, and not its physical reality. The use of additional data, such as readily available star charts (to identify the spatial clustering of member stars), is highly advisable in the absence of multi-colour photometry or large-scale radial velocity data.

\begin{acknowledgements}
      Part of this work was supported by the Brazilian agencies FAPESP and CAPES and the Brazilian-French cooperation agreement CAPES-COFECUB. This research has made use of Aladin \citep{Aladin}, R \citep{R2008}, Vizier \citep{Vizier}, and the WEBDA database \citep{WEBDA}. The authors also wish to acknowledge the valuable comments of Prof. Sekhon (U.C. Berkeley) and Prof. Mebane (U. Michigan) regarding the error estimation in the optimization procedure.
\end{acknowledgements}

\bibliographystyle{aa}
\bibliography{13881references}

\longtabL{4}{
\begin{landscape}
\begin{longtable}{lrrrrrrrrrr}
\caption{\label{table:pdfparameters}The results obtained for all the extracted fields in the PM2000 zone. The {\it c} subscripts indicate the parameters for the clusters, while the {\it f} subscripts indicate the fields ones. The correlation coefficient ($\rho_f$), mixture proportion ($n_c$) and estimated number of members ($n_{memb}$) are adimensionals, whereas all the other parameters are in mas yr$^{-1}$ units. The "Class" column is the same as in Table \ref{table:1}. The solutions are divided following their visual classifications.}\\
\hline\hline
Cluster & 
$\mu_{\alpha,c}$ & $\mu_{\delta,c}$ & 
$\mu_{\alpha,f}$  & $\mu_{\delta,f}$ &  
$\sigma_{\mu_{\alpha, f}}$ & $\sigma_{\mu_{\delta, f}}$ &
$\rho_f$ & $n_c$ & $n_{memb}$&Class\\
\hline
\endfirsthead
\caption{continued.}\\
\hline\hline
Cluster & 
$\mu_{\alpha,c}$ & $\mu_{\delta,c}$ & 
$\mu_{\alpha,f}$  & $\mu_{\delta,f}$ &  
$\sigma_{\mu_{\alpha, f}}$ & $\sigma_{\mu_{\delta, f}}$ &
$\rho_f$ & $n_c$ & $n_{memb}$&Class\\
\hline
\endhead
\hline
\endfoot
\multicolumn{11}{c}{Good} \\ \hline
NGC 2682 &$   -8.23\pm0.05 $&$   -5.72\pm0.05 $&$  -4.15\pm0.30 $&$   -5.84\pm0.25 $&$   7.20\pm0.22 $&$   6.29\pm0.20 $&$  -0.17\pm0.04 $&$   0.40\pm0.02$      &        502 & A\\
Berkeley 82 &$    3.02\pm0.61 $&$   -1.64\pm0.48 $&$  -0.13\pm0.73 $&$   -2.15\pm1.24 $&$   1.00\pm 0.00 $&$   3.50\pm1.47 $&$   0.31\pm0.22 $&$   0.33\pm0.19$   &         16 & B\\
NGC 1817 &$    2.66\pm0.18 $&$   -3.72\pm0.19 $&$   3.56\pm0.56 $&$   -6.70\pm0.67 $&$   7.16\pm0.49 $&$   8.59\pm0.54 $&$  -0.09\pm0.06 $&$   0.72\pm0.03$      &        836 & B\\
NGC 2169 &$   -2.72\pm0.40 $&$   -4.34\pm0.41 $&$  -3.64\pm3.58 $&$  -11.46\pm4.61 $&$   9.30\pm2.69 $&$  10.75\pm3.14 $&$  -0.44\pm0.29 $&$   0.86\pm0.06$      &         90 & B\\
NGC 2194 &$    0.07\pm0.20 $&$   -2.69\pm0.29 $&$  -0.31\pm0.26 $&$   -3.60\pm0.33 $&$   3.74\pm0.20 $&$   4.79\pm0.25 $&$  -0.08\pm0.06 $&$   0.21\pm0.04$      &         65 & B\\
NGC 2304 &$   -1.78\pm0.66 $&$   -5.77\pm0.68 $&$   0.89\pm1.28 $&$   -2.41\pm1.51 $&$   3.54\pm1.23 $&$   4.91\pm1.31 $&$  -0.52\pm0.19 $&$   0.64\pm0.13$      &         78 & B\\
NGC 2355 &$   -0.39\pm0.32 $&$   -4.78\pm0.30 $&$   1.03\pm0.98 $&$   -7.73\pm0.91 $&$   7.73\pm0.79 $&$   6.75\pm0.71 $&$  -0.22\pm0.10 $&$   0.64\pm0.06$      &        213 & B\\
\hline
\multicolumn{11}{c}{Intermediate} \\ 
\hline
Chupina 1 &$  -10.54\pm0.26 $&$   -3.52\pm0.21 $&$  -3.59\pm1.02 $&$   -3.17\pm0.95 $&$   5.76\pm0.71 $&$   5.75\pm0.71 $&$  -0.09\pm0.16 $&$   0.30\pm0.07$     &         22 & B\\
Chupina 2 &$    5.84\pm0.59 $&$  -12.50\pm0.69 $&$  -8.58\pm1.09 $&$   -6.68\pm1.19 $&$   3.61\pm0.73 $&$   3.96\pm1.05 $&$  -0.78\pm0.10 $&$   0.13\pm0.09$     &          3 & B\\
Chupina 3 &$   -7.32\pm0.25 $&$   -4.37\pm0.22 $&$   3.18\pm2.48 $&$   -6.55\pm2.35 $&$   5.06\pm2.05 $&$   5.90\pm1.65 $&$  -0.38\pm0.32 $&$   0.65\pm0.12$     &         17 & B\\
Chupina 4 &$   -9.91\pm0.37 $&$   -5.05\pm0.30 $&$  -2.06\pm1.11 $&$   -5.13\pm1.37 $&$   5.73\pm0.81 $&$   7.86\pm1.04 $&$  -0.27\pm0.15 $&$   0.23\pm0.08$     &         12 & B\\
Chupina 5 &$   -8.68\pm0.35 $&$   -6.90\pm0.31 $&$  -3.12\pm1.19 $&$   -8.28\pm1.11 $&$   5.05\pm0.81 $&$   5.34\pm0.87 $&$   0.07\pm0.19 $&$   0.32\pm0.11$     &         15 & B\\
NGC 7036 &$   -5.04\pm0.41 $&$   -6.84\pm0.35 $&$   0.06\pm1.26 $&$   -8.15\pm1.11 $&$   8.23\pm0.98 $&$   7.33\pm0.93 $&$   0.56\pm0.09 $&$   0.26\pm0.08$      &         20 & B\\
Roslund 1 &$   -0.06\pm0.52 $&$   -5.09\pm0.55 $&$  -1.74\pm1.06 $&$   -7.22\pm1.17 $&$   7.18\pm0.89 $&$   8.12\pm0.99 $&$  -0.08\pm0.12 $&$   0.47\pm0.08$     &         70 & B\\
Berkeley 43 &$    0.30\pm0.78 $&$   -1.81\pm0.68 $&$  -0.36\pm1.02 $&$   -3.52\pm1.65 $&$   4.38\pm1.02 $&$   8.13\pm1.31 $&$   0.08\pm0.16 $&$   0.36\pm0.15$   &         23 & C\\
Berkeley 45 &$   -1.01\pm0.96 $&$   -9.27\pm0.94 $&$   6.28\pm1.13 $&$   -3.94\pm1.39 $&$   2.44\pm1.17 $&$   3.75\pm1.10 $&$  -0.37\pm0.33 $&$   0.47\pm0.12$   &         42 & C\\
Berkeley 47 &$    0.90\pm0.90 $&$   -4.14\pm0.86 $&$   1.97\pm1.93 $&$   -4.93\pm2.46 $&$   6.79\pm1.63 $&$   9.65\pm2.05 $&$   0.65\pm0.14 $&$   0.60\pm0.12$   &         35 & C\\
King 26 &$    3.80\pm0.58 $&$   -5.72\pm0.47 $&$   1.18\pm1.99 $&$   -6.31\pm1.73 $&$   7.68\pm1.72 $&$   6.50\pm1.58 $&$   0.04\pm0.22 $&$   0.45\pm0.15$       &         21 & C\\
Skiff J0614+129 &$   -0.31\pm0.29 $&$   -2.73\pm0.37 $&$  -0.57\pm0.44 $&$   -4.31\pm0.63 $&$   4.48\pm0.32 $&$   6.56\pm0.47 $&$  -0.05\pm0.09 $&$   0.18\pm0.05$ &       29 & C\\
NGC 1807 &$    2.06\pm0.23 $&$   -4.56\pm0.24 $&$   3.57\pm0.68 $&$   -7.83\pm0.75 $&$   7.80\pm0.57 $&$   8.35\pm0.59 $&$  -0.08\pm0.07 $&$   0.66\pm0.03$      &        529 & D\\
NGC 2678 &$   -1.94\pm0.32 $&$   -2.00\pm0.28 $&$  -4.18\pm0.75 $&$   -5.34\pm0.57 $&$   8.29\pm0.55 $&$   6.10\pm0.43 $&$  -0.03\pm0.09 $&$   0.11\pm0.04$      &         16 & D\\
Teutsch 11 &$   -2.24\pm0.91 $&$   -2.61\pm0.87 $&$   0.30\pm1.28 $&$   -4.03\pm1.41 $&$   5.60\pm0.97 $&$   6.66\pm1.13 $&$   0.34\pm0.17 $&$   0.25\pm0.14$    &         16 & D\\
\hline
\multicolumn{11}{c}{Poor} \\ 
\hline
Berkeley 29 &$    0.56\pm0.45 $&$   -4.29\pm0.32 $&$   0.12\pm0.96 $&$   -6.46\pm0.88 $&$   7.43\pm0.79 $&$   6.26\pm0.68 $&$  -0.32\pm0.10 $&$   0.53\pm0.07$   &        137 & B\\
NGC 1663 &$    1.13\pm0.60 $&$   -1.32\pm0.72 $&$   3.97\pm0.65 $&$   -4.94\pm0.72 $&$   6.83\pm0.51 $&$   7.27\pm0.55 $&$  -0.14\pm0.08 $&$   0.23\pm0.07$      &         49 & B\\
NGC 2395 &$    1.22\pm0.38 $&$   -1.96\pm0.27 $&$   0.01\pm0.40 $&$   -4.32\pm0.48 $&$   5.80\pm0.31 $&$   7.17\pm0.36 $&$   0.01\pm0.06 $&$   0.27\pm0.05$      &        142 & B\\
Alessi 57 &$   -0.92\pm0.80 $&$   -7.89\pm0.68 $&$   1.11\pm1.36 $&$   -4.85\pm1.94 $&$   5.81\pm1.22 $&$   9.15\pm1.55 $&$  -0.01\pm0.19 $&$   0.30\pm0.14$     &         14 & C\\
Dias 8 &$   -4.51\pm0.48 $&$   -2.85\pm0.44 $&$  -1.13\pm1.62 $&$   -4.26\pm1.69 $&$   8.29\pm1.24 $&$   9.02\pm1.32 $&$   0.26\pm0.16 $&$   0.40\pm0.10$        &         30 & C\\
Ivanov 2 &$   -1.02\pm0.73 $&$    0.41\pm0.89 $&$  -0.89\pm0.98 $&$   -3.62\pm1.33 $&$   3.29\pm0.76 $&$   4.56\pm0.88 $&$  -0.46\pm0.21 $&$   0.22\pm0.18$      &         15 & C\\
NGC 7772 &$   17.50\pm0.57 $&$  -10.00\pm0.57 $&$   4.83\pm1.85 $&$   -8.49\pm1.63 $&$   7.59\pm1.33 $&$   6.63\pm1.21 $&$   0.13\pm0.23 $&$   0.09\pm0.06$      &          2 & C\\
DolDzim 2 &$    1.92\pm0.49 $&$   -3.25\pm0.59 $&$   3.30\pm0.92 $&$   -5.57\pm1.13 $&$   7.23\pm0.75 $&$   8.79\pm0.86 $&$  -0.05\pm0.11 $&$   0.48\pm0.07$     &         85 & D\\
DolDzim 7 &$   18.77\pm0.75 $&$  -12.13\pm0.70 $&$  -1.82\pm1.02 $&$   -5.70\pm1.23 $&$   5.79\pm0.80 $&$   7.56\pm0.94 $&$  -0.14\pm0.15 $&$   0.03\pm0.03$     &          1 & D\\
Dolidze 26 &$   -0.64\pm0.18 $&$   -2.24\pm0.25 $&$  -0.96\pm0.21 $&$   -4.13\pm0.24 $&$   6.17\pm0.17 $&$   6.84\pm0.19 $&$  -0.13\pm0.03 $&$   0.30\pm0.02$    &        536 & D\\
Dolidze 35 &$    3.87\pm0.36 $&$    0.85\pm0.37 $&$   0.32\pm0.78 $&$   -5.61\pm0.77 $&$   7.11\pm0.57 $&$   6.22\pm0.56 $&$   0.23\pm0.09 $&$   0.18\pm0.06$    &         28 & D\\
Juchert 1 &$   -0.61\pm0.69 $&$    0.95\pm0.70 $&$   4.88\pm0.77 $&$    2.17\pm1.50 $&$   1.46\pm0.72 $&$   5.75\pm1.25 $&$   0.41\pm0.17 $&$   0.17\pm0.13$     &          4 & D\\
Kronberger 13 &$    5.79\pm1.57 $&$   -5.39\pm1.57 $&$  -7.12\pm5.16 $&$    4.26\pm4.65 $&$   8.24\pm3.87 $&$   7.52\pm3.68 $&$   0.53\pm0.43 $&$   0.70\pm0.15$ &          9 & D\\
NGC 2224 &$    0.66\pm0.22 $&$   -0.91\pm0.16 $&$   0.63\pm0.16 $&$   -3.16\pm0.20 $&$   4.24\pm0.13 $&$   5.47\pm0.15 $&$  -0.14\pm0.03 $&$   0.19\pm0.02$      &        252 & D\\
NGC 2234 &$    0.14\pm0.07 $&$   -3.65\pm0.07 $&$   0.40\pm0.16 $&$   -5.88\pm0.18 $&$   5.48\pm0.13 $&$   6.11\pm0.14 $&$  -0.06\pm0.03 $&$   0.36\pm0.02$      &       2221 & D\\
NGC 2265 &$    1.60\pm0.16 $&$   -2.27\pm0.19 $&$   1.26\pm0.31 $&$   -3.82\pm0.36 $&$   4.61\pm0.25 $&$   5.24\pm0.28 $&$  -0.12\pm0.06 $&$   0.31\pm0.04$      &        149 & D\\
NGC 6525 &$   -2.39\pm0.51 $&$   -3.05\pm0.49 $&$   0.00\pm0.55 $&$   -2.63\pm0.75 $&$   5.56\pm0.45 $&$   8.24\pm0.61 $&$   0.06\pm0.08 $&$   0.21\pm0.06$      &         41 & D\\
NGC 6738 &$    1.51\pm0.17 $&$   -2.69\pm0.17 $&$  -0.36\pm0.23 $&$   -3.93\pm0.24 $&$   6.44\pm0.19 $&$   6.71\pm0.20 $&$   0.22\pm0.03 $&$   0.18\pm0.02$      &        255 & D\\
NGC 6837 &$  -23.18\pm2.77 $&$   -4.49\pm2.88 $&$  -0.13\pm0.66 $&$   -3.63\pm0.57 $&$   5.60\pm0.55 $&$   4.48\pm0.50 $&$   0.39\pm0.09 $&$   0.02\pm0.01$      &          5 & D\\
NGC 6839 &$    0.11\pm0.66 $&$   -4.88\pm0.57 $&$  -0.05\pm0.57 $&$   -6.04\pm0.54 $&$   7.48\pm0.48 $&$   7.12\pm0.46 $&$   0.12\pm0.06 $&$   0.29\pm0.05$      &        116 & D\\
NGC 6840 &$    1.08\pm0.61 $&$   -2.49\pm0.58 $&$   0.24\pm0.47 $&$   -6.07\pm0.51 $&$   6.20\pm0.41 $&$   6.34\pm0.40 $&$   0.06\pm0.06 $&$   0.24\pm0.06$      &         91 & D\\
NGC 6843 &$    3.26\pm0.68 $&$   -1.61\pm0.66 $&$   0.70\pm0.51 $&$   -4.69\pm0.55 $&$   5.72\pm0.44 $&$   6.24\pm0.46 $&$   0.13\pm0.07 $&$   0.14\pm0.07$      &         33 & D\\
NGC 6858 &$   -0.27\pm0.43 $&$   -3.35\pm0.34 $&$   0.46\pm0.42 $&$   -4.43\pm0.43 $&$   7.27\pm0.33 $&$   7.71\pm0.35 $&$   0.14\pm0.05 $&$   0.36\pm0.04$      &        261 & D\\
NGC 6950 &$   -0.89\pm0.34 $&$   -6.36\pm0.39 $&$  -0.01\pm0.27 $&$   -6.38\pm0.27 $&$   6.44\pm0.21 $&$   6.72\pm0.22 $&$   0.42\pm0.03 $&$   0.09\pm0.03$      &         80 & D\\
NGC 7084 &$   -5.50\pm0.51 $&$  -12.09\pm0.52 $&$  -0.04\pm0.47 $&$   -9.43\pm0.43 $&$   7.43\pm0.32 $&$   7.31\pm0.33 $&$   0.29\pm0.05 $&$   0.19\pm0.04$      &         97 & D\\
Riddle 15 &$    2.73\pm0.90 $&$   -8.10\pm0.88 $&$  -2.42\pm3.10 $&$   -2.21\pm4.75 $&$   7.27\pm2.27 $&$  11.87\pm3.67 $&$   0.76\pm0.18 $&$   0.24\pm0.15$     &          3 & D\\
Teutsch 12 &$    0.81\pm0.37 $&$   -1.98\pm0.35 $&$  -0.60\pm0.79 $&$   -2.89\pm0.96 $&$   4.97\pm0.63 $&$   6.22\pm0.78 $&$  -0.08\pm0.13 $&$   0.31\pm0.09$    &         29 & D\\
\hline
\end{longtable}
\end{landscape}
}

\begin{landscape}
\begin{table}
\caption{Example of the membership probability table for the remnant NGC 7772. The PM2000 and 2MASS columns points to the identifiers of the object in the respective catalogues. The Prob. column indicates the membership probability as calculated from eq. \ref{eq:prob}.} 
\label{table:example}
\centering
\begin{tabular}{l r r r r r r r r r r r r r r r r r}
\hline\hline
Cluster & PM2000&&&$\alpha_{J2000}$ &&&$\delta_{J2000}$ & $\sigma_\alpha$ & $\sigma_\delta$ & $\mu_\alpha\cos\delta$ & $\mu_\delta$ & $\sigma_{\mu_\alpha\cos\delta}$ & $\sigma_{\mu_\delta}$ & $V_M$ & $\sigma_{V_M}$ & 2MASS & Prob.\\
&  &h & m&  s & $^{\circ}$ & '&  '' & arcsec & arcsec & mas yr$^{-1}$ & mas yr$^{-1}$ & mas yr$^{-1}$ & mas yr$^{-1}$ & mag & mag &  & \%\\
\hline                        
NGC 7772&664742&23&51&20.0035&16&14&31.344&0.039&0.043&  2.5& -2.5&3.9&4.0&15.722&0.152&  23512000+1614312&0.0\\
NGC 7772&664906&23&51&33.7205&16&20&33.280&0.055&0.057&  4.4& -7.7&4.3&4.3&15.998&0.015&  23513371+1620334&0.3\\
NGC 7772&664928&23&51&35.4557&16&12&49.718&0.024&0.025& -1.3&-19.3&0.8&0.8&13.660&0.054&  23513545+1612494&0.0\\
NGC 7772&664949&23&51&37.3487&16&18&02.019&0.049&0.043&  5.9& -7.7&3.9&3.9&15.543&0.117&  23513734+1618020&0.4\\
NGC 7772&664966&23&51&38.9136&16&12&22.572&0.029&0.030&  2.1&  3.0&2.1&2.1&10.357&0.050&  23513891+1612224&0.0\\
NGC 7772&664970&23&51&39.1210&16&18&13.397&0.030&0.031&-15.1&-37.1&3.7&3.7&13.266&0.065&  23513911+1618132&0.0\\
NGC 7772&664998&23&51&41.3285&16&10&43.795&0.034&0.035& -6.4&-16.8&3.7&3.7&15.432&0.136&  23514131+1610436&0.0\\
NGC 7772&665026&23&51&42.8851&16&14&06.988&0.023&0.024& 11.2&-10.3&0.8&0.8&11.125&0.056&  23514289+1614067&0.0\\
NGC 7772&665044&23&51&44.4787&16&20&17.501&0.039&0.040&  4.8&-10.9&3.9&3.9&15.466&0.121&  23514448+1620175&0.2\\
NGC 7772&665048&23&51&44.7867&16&14&43.252&0.024&0.025& 15.6&  3.4&0.8&0.8&14.376&0.082&  23514479+1614431&0.0\\
NGC 7772&665067&23&51&46.2475&16&14&23.413&0.023&0.024& 17.4&-10.1&0.8&0.8&12.578&0.059&  23514626+1614231&97.2\\
NGC 7772&665069&23&51&46.3839&16&14&56.755&0.025&0.026& 17.6& -9.9&0.8&0.8&13.514&0.136&  23514640+1614564&97.3\\
NGC 7772&665087&23&51&47.5439&16&12&54.168&0.037&0.041&  8.7&-20.7&3.8&3.9&15.753&0.185&  23514753+1612539&0.3\\
NGC 7772&665088&23&51&47.5485&16&15&55.216&0.027&0.028& 48.8& -6.0&0.8&0.8&12.748&0.066&  23514760+1615549&0.0\\
NGC 7772&665096&23&51&48.0574&16&14&00.102&0.023&0.024&  5.2& -3.8&0.8&0.8&13.723&0.067&  23514805+1613598&0.0\\
NGC 7772&665101&23&51&48.1263&16&15&04.932&0.026&0.027& 13.4&-12.3&0.8&0.8&13.436&0.071&  23514814+1615045&0.0\\
NGC 7772&665113&23&51&48.9226&16&11&00.074&0.026&0.027& -2.0& -9.1&0.8&0.8&14.405&0.097&  23514892+1610597&0.0\\
NGC 7772&665147&23&51&52.4475&16&11&54.192&0.035&0.033& 39.7&  1.5&5.7&5.7&15.326&0.148&  23515244+1611540&1.8\\
NGC 7772&665148&23&51&52.5442&16&11&17.381&0.026&0.027& 10.9&-40.4&0.8&0.8&12.733&0.061&  23515255+1611165&0.0\\
NGC 7772&665203&23&51&57.9668&16&17&31.784&0.027&0.028& 24.2&-11.9&0.9&0.8&14.148&0.079&  23515799+1617313&0.0\\
NGC 7772&665223&23&51&59.5060&16&19&48.341&0.032&0.035& -9.3&-13.4&3.8&3.8&15.130&0.209&  23515950+1619482&0.0\\
NGC 7772&665236&23&52&00.3693&16&09&44.603&0.029&0.030&  3.3&-16.9&2.1&2.1&10.978&0.058&  23520036+1609444&0.0\\
NGC 7772&665266&23&52&03.3627&16&17&41.938&0.026&0.027&  1.2& -1.8&0.8&0.8&12.921&0.071&  23520336+1617418&0.0\\
NGC 7772&665271&23&52&03.9432&16&14&13.038&0.027&0.030& -2.1& -3.5&3.7&3.7&15.152&0.141&  23520393+1614129&0.0\\
NGC 7772&665273&23&52&04.3616&16&19&18.151&0.039&0.042&  0.1& -7.9&3.9&4.0&15.670&0.212&  23520435+1619180&0.0\\
NGC 7772&665279&23&52&04.7888&16&11&32.168&0.032&0.035&  7.9& -0.3&3.7&3.7&15.342&0.096&  23520479+1611321&0.1\\
\hline                                  
\end{tabular}
\end{table}
\end{landscape}

\end{document}